\documentclass[aps,prl,epsfigure,twocolumn,superscriptaddress]{revtex4-2}
\usepackage[colorlinks=true,linkcolor=blue,urlcolor=blue,citecolor=blue,pdfusetitle]{hyperref}
\usepackage[utf8]{inputenc}
\usepackage[english]{babel}
\usepackage{amsmath}
\usepackage{graphicx,epstopdf}
\usepackage[table,xcdraw]{xcolor}
\usepackage{lipsum}
\usepackage{amsfonts}
\usepackage{bbm}
\usepackage{amssymb}
\usepackage{enumerate}
\usepackage{color}
\usepackage{xcolor}
\usepackage{physics}
\usepackage{times,txfonts}
\usepackage{orcidlink}
\usepackage[normalem]{ulem}

\graphicspath{{figures/}{old_figures/}}

\begin{document}

\title{Coherent Control of Photon Correlations in Trapped Ion Crystals}

\author{K. Singh}
\affiliation{Department of Optics, Palack\'{y} University, 17. listopadu 12, 771 46 Olomouc, Czech Republic}

\author{A. Cidrim}
\affiliation{Departamento de F\'isica, Universidade Federal de S\~ao Carlos, Rodovia Washington Lu\'is, km 235—SP-310, 13565-905 S\~ao Carlos, SP, Brazil}

\author{A. Kovalenko}
\affiliation{Department of Optics, Palack\'{y} University, 17. listopadu 12, 771 46 Olomouc, Czech Republic}

\author{T. Pham}
\affiliation{Institute of Scientific Instruments of the Czech Academy of Sciences, Kr\'{a}lovopolsk\'{a} 147, 612 64 Brno, Czech Republic}

\author{O. Číp}
\affiliation{Institute of Scientific Instruments of the Czech Academy of Sciences, Kr\'{a}lovopolsk\'{a} 147, 612 64 Brno, Czech Republic}

\author{L. Slodi\v{c}ka}
\affiliation{Department of Optics, Palack\'{y} University, 17. listopadu 12, 771 46 Olomouc, Czech Republic}

\author{R. Bachelard}
\affiliation{Departamento de F\'isica, Universidade Federal de S\~ao Carlos, Rodovia Washington Lu\'is, km 235—SP-310, 13565-905 S\~ao Carlos, SP, Brazil}

\date{\today}

\begin{abstract}
While the spontaneous emission from independent emitters provides spatially uncorrelated photons --- a typical manifestation of quantum randomness, the interference of the coherent scattering leads to a well-defined intensity pattern --- a feature described by linear optics. We here demonstrate experimentally how the interplay between the two mechanisms in large systems of quantum emitters leads to spatial variations of photon correlations. The implementation with trapped ion crystals in free space allows us to observe the anti-correlation between photon rates and variance of the photon number distributions in chains of up to $18$~ions. For smaller crystals of four ions, the transition from antibunching to bunching and super-Poissonian statistics of the scattered light is reported. For higher numbers of scatterers, the photon statistics still display a strong deviation from the fully incoherent scattering case. Our results illustrate how the interference of coherent scattering, combined with spontaneous emission, provides a control mechanism for the light statistics.
\end{abstract}

\maketitle

{\em Introduction---}Intensity correlations were first introduced as a technique to measure the angular size of stars~\cite{Hanbury1952,Hanbury1954,HBT:1956b}. Yet the puzzling observation of bunching from sources otherwise expected to be chaotic, combined with the ideas that two different photons cannot interfere~\cite{Dirac1930}, stimulated the birth of quantum optics: Light is characterized by the photon autocorrelation functions at all orders~\cite{Glauber:1963a,Glauber:1963b,Glauber:1963c}, and field-field correlations, which are associated with the optical spectrum~\cite{Wiener1930,Khintchine1934}, contain information which is different from intensity correlations. Nowadays, two-photon correlations are commonly used, for instance, to differentiate coherent light (e.g., laser light, composed of completely uncorrelated photons) from chaotic sources (characterized by photon bunching) and single-photon emitters (producing isolated photons). The latter typically exhibit sub-Poissonian statistics [$g^{(2)}(0)<1$] and photon antibunching [$g^{(2)}(0)<g^{(2)}(\tau)$], which are non-classical features~\cite{Kimble1977,Dagenais1978,Loudon:book,Scully1997}. Apart from its fundamental interest to explore quantum optics, antibunching has become a key feature of single-photon sources in the context of quantum technologies. It is particularly important for quantum cryptography protocols~\cite{Beveratos2002}, and solid state quantum emitters such as quantum dots~\cite{Michler2000,Bozzio2022}, and more recently two-dimensional materials~\cite{Gao2023}, are alternative sources of antibunched photons.

Interestingly, the light statistics of single emitters can themselves be manipulated, for example, using sequential pulses to generate photon-number entangled states~\cite{Wein2022}. More fundamentally, the antibunching feature of driven single two-level emitters can be interpreted as a destructive interference between coherently and incoherently scattered components~\cite{Dalibard1983}. As a consequence, filtering one of the two components allows one to select either the properties of the coherently-scattered drive, or the bunched photons from the higher-order incoherent component~\cite{delValle2012, LopezCarreno2018, ZubizarretaCasalengua2020, Hanschke2020, Phillips2020, Masters2023}. In principle, the radiation from several emitters can then be made to interfere, changing the balance between single and two-photon components and modifying the statistics of the resulting light~\cite{Bojer2024}. This control feature offered by the many-emitter case has been demonstrated for several cold atoms weakly coupled to a nanofiber in a one-dimensional, chiral geometry~\cite{Cordier2023}. In free space, this controllable balance has only been demonstrated so far for pairs of ions~\cite{Wolf2020}, leaving open questions on its scalability and the thermodynamic limit of the phenomenon for a larger number of emitters.

\begin{figure}[t!]
\centering	
\includegraphics[width=\columnwidth]{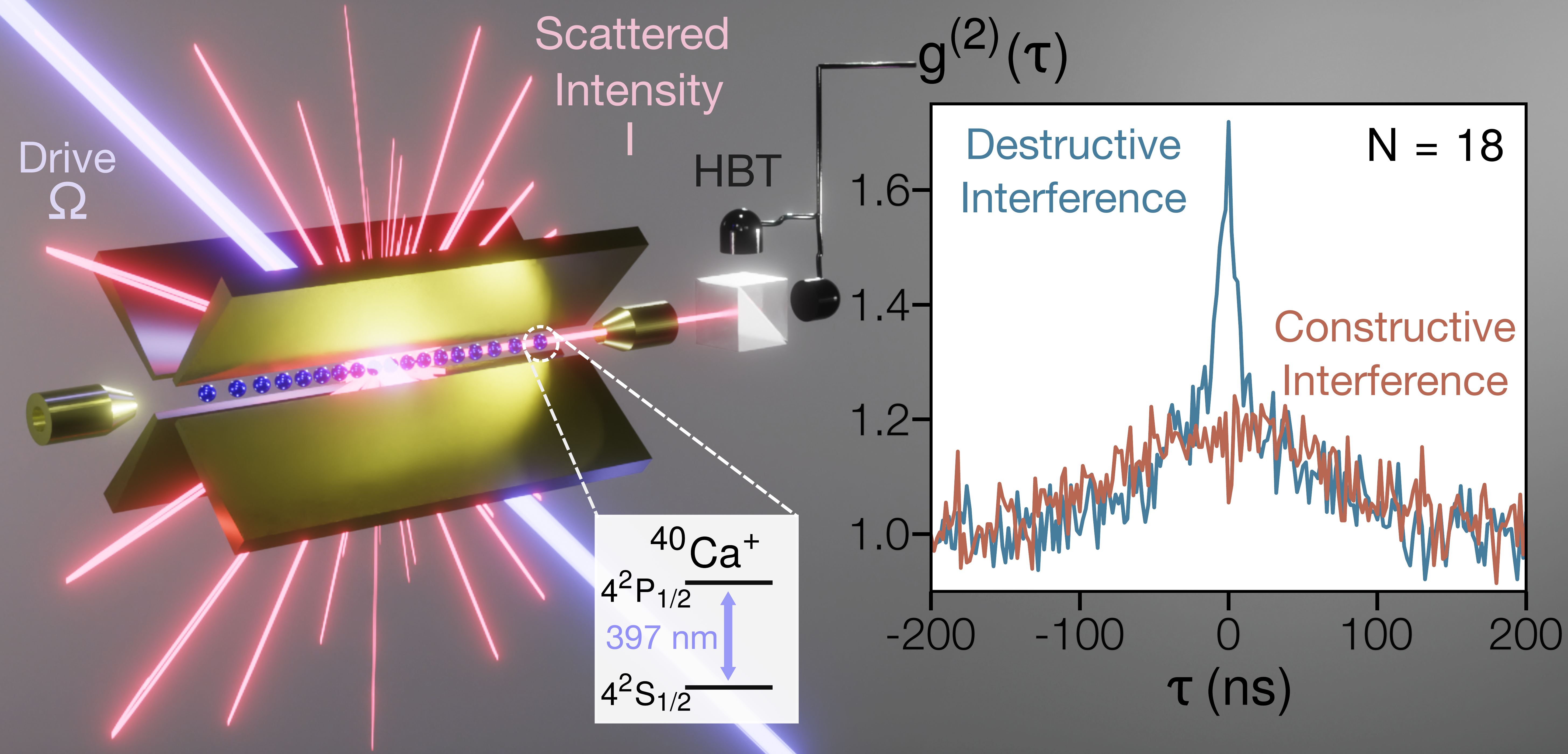}
\caption{Schematic description of the experiment, where the light scattered by a long chain ($N=18$) of non-interacting $^{40}\text{Ca}^+$ ions is monitored through a Hanbury-Brown Twiss (HBT) setup. The weakly driven ($s=0.6$) two-level emitters present an interference pattern of the intensity $I$ from the elastically scattered light, which is anti-correlated to the two-photon correlations signal $g^{(2)}(0)$.}
\label{fig:scheme}
\end{figure}

In this work, we turn to mesoscopic ensembles of quantum emitters in free space, showing how photon-photon correlations can be manipulated using the interference pattern of coherent scattering. More specifically, two-photon correlations from weakly-driven ion crystals are shown to be anticorrelated with the intensity pattern, even for an intermediate number of independent emitters --- see Fig.~\ref{fig:scheme}. We use a tunable one-dimensional chain of up to $N=18$ trapped ions with an excitation geometry which provides an efficient collection of the light: Tuning the inter-ion distance allows us to monitor destructive and constructive interference, which are associated with maxima and minima of intensity correlations, respectively. Furthermore, the second-order photon autocorrelation differs from the value of a Gaussian field even in the mesoscopic limit. Our work thus paves the way for the controllable production of quantum light in large systems of quantum emitters.

{\em Linear ion crystal setup---}We here implement a free-space paraxial detection of the light scattered from $^{40}$Ca${^+}$ ion strings in a linear Paul trap of radial confinement of $\omega_x \approx \omega_y \approx (2\pi) 2.0$ MHz and tunable axial frequency of up to $\omega_z\approx (2\pi) 1.1$ MHz. The configuration, as depicted in Fig.~\ref{fig:scheme}, allows us to reach an efficient collection of coherently scattered photons along the linear ion chain axis with intrinsic spatial indistinguishability of the contributing ions and thus a high interference visibility~\cite{wolf2016visibility,obvsil2018nonclassical,kovalenko2023emergence}. The scattered light is here probed for ion numbers ranging from $N=2$ to $N=18$. Although the coupling efficiency to the detection optical mode remains constant for up to hundreds of ions, this upper limit for the ion number is set by the requirement of near-equal excitation by the 397~nm (Gaussian) excitation beam, whose Rabi frequency $\Omega$ and detuning $\Delta$ define a close-to-equal saturation parameter $s=2\Omega^2/(4\Delta^2+\Gamma^2)$ for each ion, with $\Gamma=(2\pi) 21.2$~MHz the natural linewidth of the $4^2{\rm S}_{1/2}\leftrightarrow 4^2{\rm P}_{1/2}$ transition. The large average inter-ion distance (on the order of a few $\mu$m, set by the typical frequency-dependent Coulomb length scale $\ell_0=(e^2/4\pi\epsilon_0 M\omega_z^2)^{1/3}$, with $M$ the ion mass, $e$ the electron charge, and $\epsilon_0$ the free-space permittivity \cite{James1998}) guarantees the mutual independence of ions for the optical dipole transition~\cite{devoe1996observation}, and it is controlled through the tunable setting of axial trapping potential $U_{\rm tip}$ applied to hollow-tip electrodes. Varying the inter-ion distances effectively allows us to continuously change the interference pattern in the monitored direction from destructive to constructive points. This in turn alters the observed photon statistics, as demonstrated in the plots of the second-order autocorrelation functions $g^{(2)}(0)$ in Fig.~\ref{fig:scheme}. The light scattered from the ion chain is measured using a Hanbury-Brown and Twiss (HBT) setup with two single-photon avalanche diodes (SPADs). The photon rates for different axial trapping potentials provide the interference patterns emerging from the coherent scattering by many ions, while the photon correlations are deduced from the time-tagged measurements on the two SPADs. The number of ions $N$ in the trap is determined by ion counting in the transversal direction using an EM-CCD camera --- see~\cite{SM} for further details.

{\em Light statistics from a chain of quantum emitters---}The emission from the driven linear ion crystal is modeled by considering an ensemble of $N$ non-interacting two-level systems. The corresponding far-field electric field is given by $\hat{E}^+\propto\sum_j e^{-ik\hat{n}\cdot\mathbf{r}_j}\hat{\sigma}^-_j$, with $k$ the transition wavenumber and $\hat{\sigma}_j^-$ ($\hat{\sigma}_j^+$) the lowering (raising) operators of the Pauli algebra. The ions' positions $\mathbf{r}_j$ in the one-dimensional crystal are calculated from the equilibrium points resulting from the competing harmonic confinement of the Paul trap and the inter-ion Coulomb repulsion \cite{James1998}. The dependence on the direction $\hat{n}$ of emission is hereafter kept implicit. In the steady state, in which the experiment is operated, the normalized second-order correlation function can be expressed as
\begin{equation}
g^{(2)}(\tau)=\frac{\langle \hat{E}^-(t)\hat{E}^-(t+\tau)\hat{E}^+(t+\tau)\hat{E}^+(t) \rangle}{I^2},    
\end{equation}
with $I=\langle\hat{E}^-(t)\hat{E}^+(t)\rangle$ the intensity. While in the analytical derivation $\langle\cdot\rangle$ corresponds to the expectation value, in the experiment it is accessed by a time average.

The difference in two-photon correlations $g^{(2)}(0)$ between the constructive and destructive interference directions for intensity can be appreciated in Fig.~\ref{fig:scheme}, where the temporal profiles $g^{(2)}(\tau)$ at local maximum and minimum of the intensity are presented for a chain of $N=18$ ions and $s=0.6$. While at the destructive interference the light exhibits a substantial photon bunching, $g^{(2)}(0)\approx 1.75$, at the constructive one the zero-delay value reaches only $g^{(2)}(0)\lesssim 1.20$. 

As the contributing emitters are many wavelengths apart, they can be treated as independent, and we will now see that the observed features stem from the interplay between interference of coherently scattered photons and spontaneous emission. Driven by a resonant plane wave of wavevector $\mathbf{k}_L$, the equal-time two-photon correlation $g^{(2)}(0)$ for a set of $N$ independent two-level emitters is given by~\cite{SM}
\begin{align} 
g^{(2)}(0)=\frac{\left(2-\frac{2}{N}+\frac{4}{Ns}\right)I_\textrm{SE}^2 +4\left(1-\frac{2}{N}\right)I_\textrm{SE}I_\textrm{coh}+|E_\textrm{coh}^2-\Phi|^2}{\left(I_\textrm{SE}+I_\textrm{coh}\right)^2}, \label{eq:g20}
\end{align} 
where we have introduced the coherently scattered field $E_\textrm{coh}=\frac{\sqrt{s/2}}{1+s}\sum_j e^{-i(k\hat n-\mathbf{k}_L)\cdot \mathbf{r}_j}$ and intensity $I_\textrm{coh}=|E_\textrm{coh}|^2$ --- it can be thought as the single-photon component of the field, which produces interference. The spontaneously emitted intensity, $I_\textrm{SE}=Ns^2/2(1+s)^2$, is isotropic and this multi-photon component of the field dominates the emission in the destructive interference direction where the coherent scattering cancels out. The term $\Phi=\frac{s/2}{(1+s)^2}\sum_j e^{-2i(k\hat n -\mathbf{k}_L)\cdot \mathbf{r}_j}$ 
stems from the two-level nature of the emitters, and from their incapacity to emit two photons at a time. Note that the use of a separable state to describe the atoms reflects that the observed features for the light statistics rely on interference from independent emitters, rather than filtering of the light from a given emitter~\cite{Schulte2015,Hanschke2020,Masters2023}, or interaction between the emitters~\cite{Hinney2021,Pizzi2023,Rosario2024}.

In the strong drive limit ($s\to \infty$), the coherent terms in Eq.~\eqref{eq:g20} become negligible and the two-photon correlations for spontaneously emitted photons from $N$ independent emitters --- or in presence of another phase-randomization mechanism such as thermal motion~\cite{Loudon:book,Lassegues2023,kovalenko2023emergence} --- is recovered
\begin{equation}
g^{(2)}(0)=2\left(1-\frac{1}{N}\right),\label{eq:g20SE}  
\end{equation}
and it is isotropic. Differently, at finite saturation parameter the coherently scattered field, whose amplitude depends on the direction of observation $\hat{n}$, leads to a spatial modulation of the second-order correlation $g^{(2)}(0)$. In particular, in the limit of elastic scattering ($s\to 0$), 
only the last terms in the numerator and denominator remain and provide
\begin{equation}
g^{(2)}(0)=\frac{|E_\textrm{coh}^2-\Phi|^2}{I_\textrm{coh}^2}.  \label{eq:g20el}
\end{equation}
Remarkably, this formula predicts antibunching, when $\Phi$ is in phase with $E_\textrm{coh}^2$, that is, $|\arg(E_\textrm{coh}^2)-\arg(\Phi)|<\pi/2$. This is for example the case in the forward scattering direction, $\hat n =\mathbf{k}_L/k$, where a strong constructive interference for intensity is obtained, and $E_\textrm{coh}^2$ and $\Phi$ both have zero phase. But Eq.~\eqref{eq:g20el} also predicts superbunching in the destructive interference directions ($I_\textrm{coh}=0$). In the latter case, including the spontaneous emission to correct the emerging divergence leads to the following prediction
\begin{equation}\label{eq:g20dest}
g^{(2)}_\textrm{dest}(0)\approx \frac{4}{Ns} +\frac{|\sum_j e^{-2i(k\hat n-\mathbf{k}_L)\cdot \mathbf{r}_j}|^2}{N^2s^2}.  
\end{equation}
Hence, the lower the saturation parameter, the stronger the superbunching. More specifically, $Ns$, which corresponds to the number of excitations in the system at low~$s$, is the control parameter to manipulate the photon correlations~\cite{Bojer2024}.

\begin{figure}[t!]
\includegraphics[width=\columnwidth]{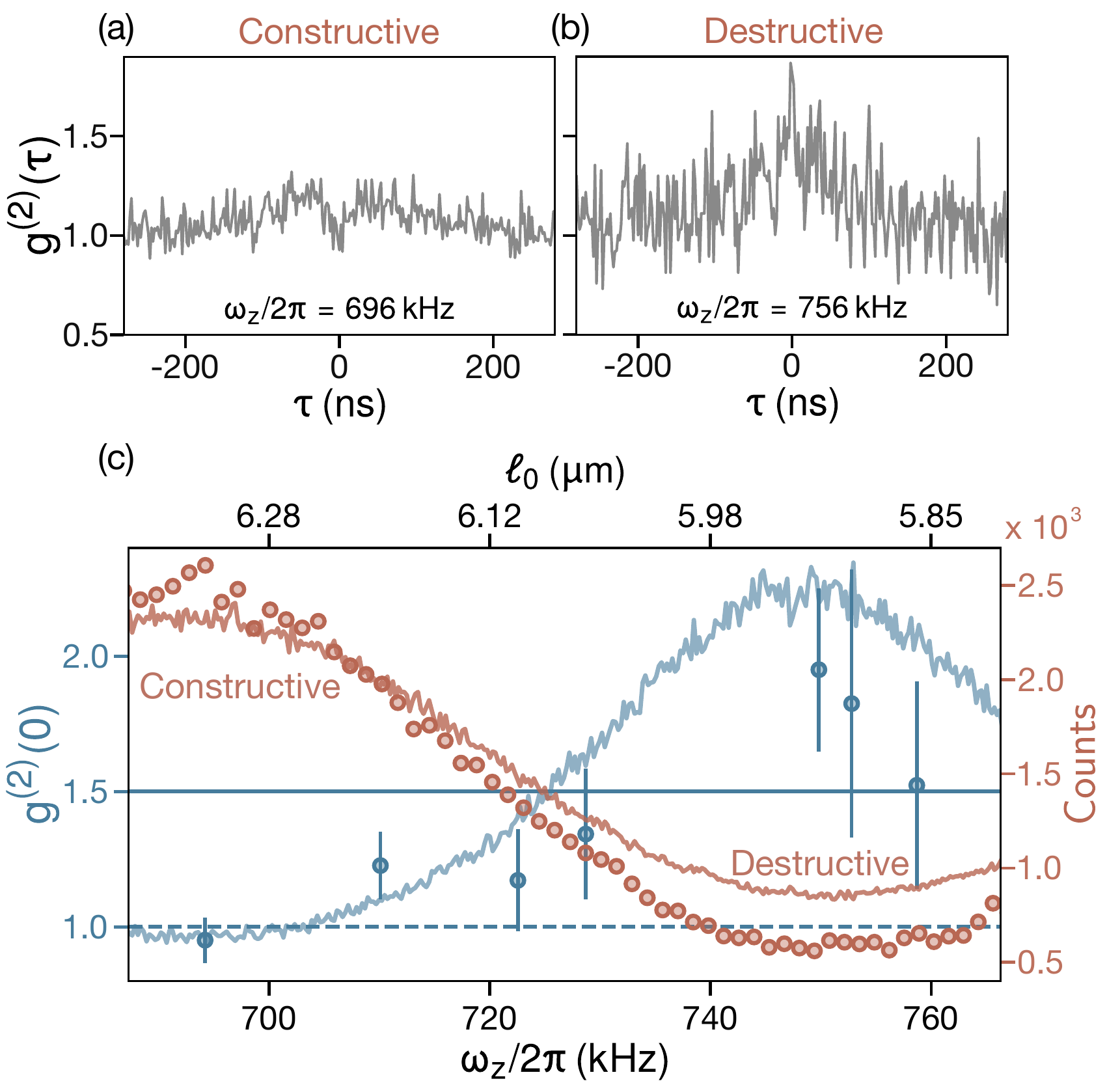}
\caption{(a-b) Experimental data on the temporal two-photon correlations for $N=4$ ions and $s=0.62$, measured in constructive and destructive interference directions. In panel (a), the dip at zero time delay, $g^{(2)}(0)<g^{(2)}(\tau)$, corresponds to antibunching. (c) Two-photon equal-time correlations $g^{(2)}(0)$, as the distance between ions is tuned to switch from constructive to destructive interference through the trap frequency $\omega_z$. The red and blue circles are the experimental points for photon counts and $g^{(2)}(0)$, respectively, while the solid lines show the corresponding theoretical predictions. For each axial trapping frequency $\omega_z$, the values were averaged over $10^3$ numerical realizations. This accounts for the jitter in the positions of the ions and for the occupation of electronic levels which do not scatter in the monitored light mode (see main text), and results in the visible fluctuations. The discrepancy between the experimental and numerical curve may be due to the ion motion, which induces timescales not accounted for by the theory. The horizontal lines stand for Poissonian light (dashed line) and the phase-randomized prediction~\eqref{eq:g20SE} (solid dark-blue line).}
\label{fig:g2inter}
\end{figure}

Let us now turn to the lower ion number $N=4$, for which larger fluctuations of the photon statistics can be observed --- $s\gtrsim 0.6$ is required in our experiment to achieve sufficient photon count rates. This anticorrelation between the extrema of the intensity and of the two-photon correlations discussed above are shown in Fig.~\ref{fig:g2inter}(a-b). Antibunching is detected in a direction of constructive interference for intensity  [$g^{(2)}_\textrm{cons}(0)\approx 0.95\pm 0.08$, see panel (a)], whereas  bunching [$g^{(2)}_\textrm{dest}(0)\approx 1.9 \pm 0.3$, see panel (b)] is measured where a destructive interference occurs. 

A systematic study on this ion chain reveals how the second-order photon autocorrelation function $g^{(2)}(0)$ is anticorrelated with the intensity for a relatively weak saturation parameter ($s= 0.6$). More specifically, it oscillates around the prediction for phase-randomized emission~\eqref{eq:g20SE}, $g^{(2)}(0)=1.5$ for this case where $N=4$. This demonstrates how coherent scattering modulates the light statistics thanks to the interference of the single-photon (coherent) component. 

The theoretical prediction~\eqref{eq:g20} for the $g^{(2)}(0)$ relies on the hypothesis of two-level emitters without any decoherence mechanism other than spontaneous emission. The simulations presented in Fig.~\ref{fig:g2inter}(c) incorporate two additional effects, namely position jittering due to residual thermal motion and non-fluorescence of the ions. We assumed that the position uncertainties of each ion follow a Gaussian distribution with RMS width of $70$~nm, which includes  radial and axial thermal position fluctuations at the Doppler cooling limit. Second, despite the large Zeeman shifts resulting from the applied magnetic field, the multilevel structure of the emitters still affects the photon statistics due to the residual population in the metastable $3^2 {\rm D}_{3/2}$ manifold and other Zeeman sub-level of the $4^2 {\rm S}_{1/2}$ state. The total population probability of these levels is estimated to $40\%$, from a measured dark-resonance spectrum of a single ion. This effect is included in the simulations by switching off the contribution of randomly sorted ions. 

\begin{figure}[t!]
\centering	
\includegraphics[width=\columnwidth]{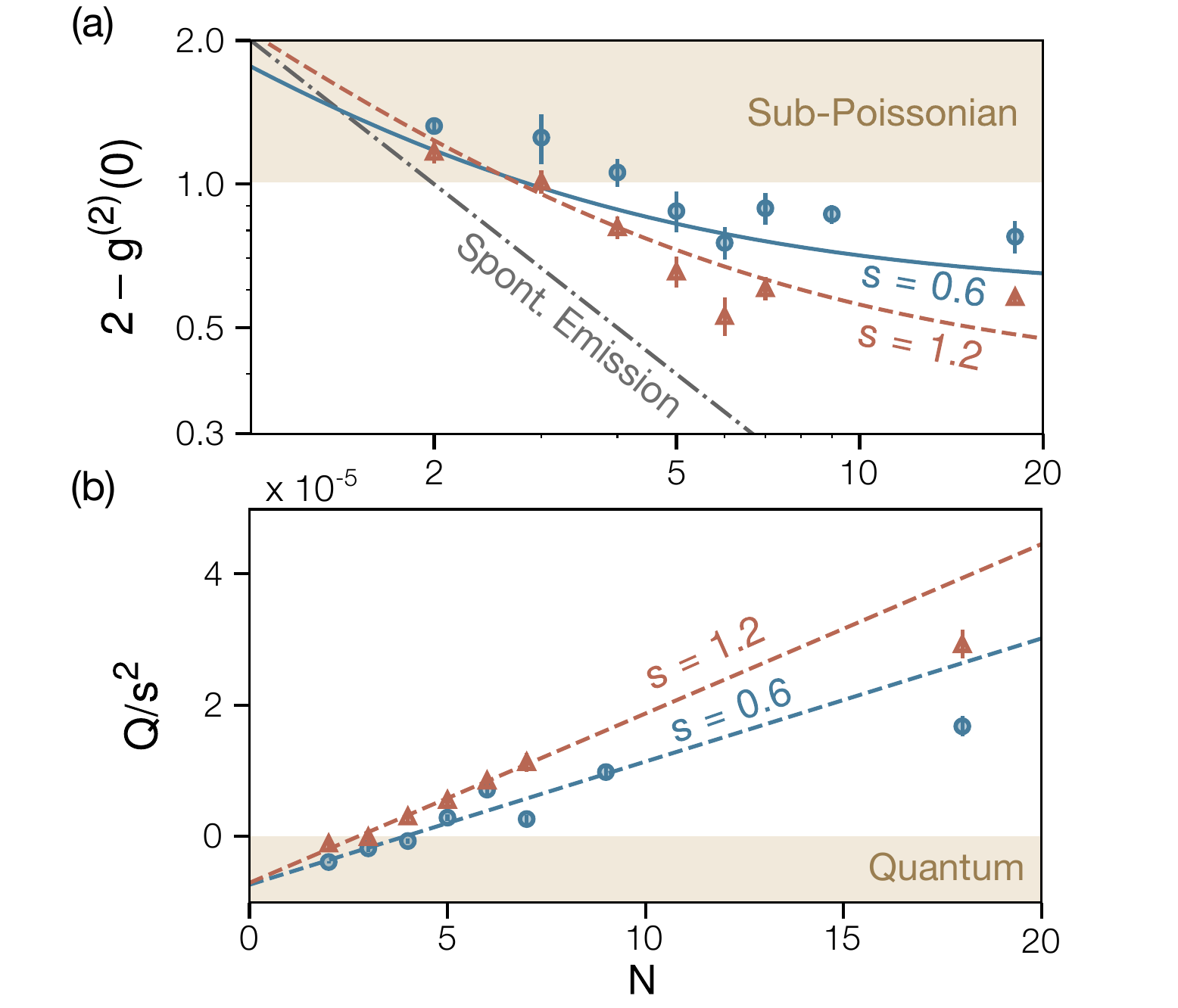}
\caption{(a) Second-order photon autocorrelation function $g^{(2)}(0)$ in the constructive interference direction, for an increasing number of ions, $N=2...18$. The blue circles and red triangles correspond to the experimental data, for $s\approx 0.6$ and $1.2$, respectively, while the lines stand for the theoretical prediction~\eqref{eq:g20cons}, for the same parameters. Values above $1$ (shaded area) correspond to sub-Poissonian statistics, $g^{(2)}(0)<1$, while the black dash-dotted line refers to the spontaneous emission prediction $2(1-1/N)$. The theoretical prediction for $g^{(2)}(0)$, based on the hypothesis of a speckle pattern and which aims at capturing the large-$N$ behaviour, does not converge to $0$ for $N=1$. (b) Mandel $Q$ parameter, normalized by the square saturation parameter $s^2$ for clarity, as a function of the particle number $N$, and for saturation parameters $s=0.6$ and $1.2$. The dashed lines correspond to linear fits over the points for $N<10$.}
\label{fig:thermo}
\end{figure}

{\em Mesoscopic regime---}Let us now discuss the two-photon correlations in the regime of a growing number of emitters. If phase-randomized emission predominates, whether it stems from spontaneous emission or other mechanisms~\cite{Loudon:book}, equation~\eqref{eq:g20SE} predicts a $1/N$ convergence toward the thermodynamic limit of thermal light, that is, $g^{(2)}(0)=2$ for $N=\infty$.

The presence of coherent scattering, however, changes this scenario, as we shall now see. We focus on the constructive interference direction, which provides access to higher photon rates and thus better photon coincidence statistics. The harmonic trap results in an irregular spacing between the ions, so the chain can be considered as a disordered system of independent scatterers. Hence its average elastic background $\overline{I}_\textrm{coh}$, treated as a speckle pattern~\cite{Goodman2020}, scales as $sN/2(1+s)^2$. We then assume that the maxima of intensity scale as twice the background, $I_\textrm{coh}^\textrm{cons}=sN/(1+s)^2$, as expected from a fully developed speckle. Furthermore, the $\Phi$ term can be neglected compared to $E_\textrm{coh}^2$ since it corresponds to a simple (random) sum, as compared to a double sum. These approximations lead to the following prediction for the equal-time intensity correlations in constructive interference directions
\begin{equation}
g^{(2)}_\textrm{cons}(0)\approx 2-\frac{1}{(1+s/2)^2}-\frac{2s(6+s)}{N(2+s)^2}.  \label{eq:g20cons}
\end{equation}
In this equation, the first two terms describe the balance between spontaneous emission, with the limit $g^{(2)}(0)=2$ reached at large $s$ and $N$, and elastic scattering, with the limit $g^{(2)}(0)=1$ at low $s$. Finally, the last term describes a $1/N$ correction, similarly to the phase-randomized case, see Eq.~\eqref{eq:g20SE}. 

The $g^{(2)}(0)$ measured in our trap, for ion numbers ranging from $N=2$ to $18$ and in constructive interference, is in good agreement with the prediction~\eqref{eq:g20cons}, as can be seen in the measured data corresponding to blue circles in Fig.~\ref{fig:thermo}(a). The second-order photon autocorrelation function does not converge to the chaotic light value of $2$ in the mesoscopic limit, as long as elastic scattering is present ($s<\infty$), or unless some mechanism provides phase randomization. Furthermore, if the saturation parameter is increased, here from $s= 0.6$ to $1.2$, the statistics become closer to the prediction of fully-randomized phase from the emitters. This is the signature that spontaneous emission --- with the associated phase randomization --- is progressively taking over, and the prediction~\eqref{eq:g20SE} is thus expected to be recovered in the large $s$ limit; in particular, any spatial variation of $g^{(2)}(0)$ is lost in this limit. Similarly, while Eq.~\eqref{eq:g20el} predicts antibunching at an arbitrary particle number, the conditions to detect it are increasingly stringent as the ion number is increased. The fact that the term $\Phi$, which provides a below-unity $g^{(2)}(0)$, is a simple sum, versus a double sum for $E_\textrm{coh}^2$, makes its relative contribution smaller as $N$ increases.

The transition from sub-Poissonian [$g^{(2)}(0)<1$] to super-Poissonian [$g^{(2)}(0)>1$] light statistics can be better appreciated in Fig.~\ref{fig:thermo}(b), where the Mandel parameter $Q=\langle\hat{n}\rangle \left(g^{(2)}(0)-1\right)$ is presented. On the one hand, the parameter exhibits an almost-linear behaviour with the emitter number~$N$, which takes its origin in the scaling of the average photon number $\langle\hat{n}\rangle$. On the other hand, sub-Poissonian statistics are observed until a larger ion number when the saturation parameter is smaller: It is yet another manifestation that quantum light may be produced from mesoscopic  systems of emitters, provided that phase randomization mechanisms such as spontaneous emission are kept at a minimum.

{\em Discussion and prospects---}We have thus reported on the spatial modulation of the second-order photon autocorrelation function in mesoscopic chains of up to $18$ ions, with its extrema being anticorrelated with those of the intensity. This phenomenon stems from the interference of the coherent scattering from the weakly driven quantum emitters, as well as the absence of decoherence mechanism other than spontaneous emission, thanks in particular to the strong confinement provided by the ion trap. This interference-based modulation of the light statistics is thus different from the one obtained by frequency-filtering the light from single emitters~\cite{Hanschke2020,Phillips2020}, or by considering correlated atomic states~\cite{Pizzi2023}. We point out that while we reported antibunching and superbunching for a chain of $N=4$ ions, these features can in principle be achieved in systems of arbitrary size, provided that the driving power, and more precisely $sN$, is low enough. The competition between coherent scattering and spontaneous emission thus offers a mechanism to control the photon correlations even in mesoscopic ensembles of quantum emitters, where a single sample provides access to a range of light statistics. 

Nevertheless, increasing the system size leads to several intrinsic limitations. On the one hand, the manifestations of the inherent decoherence mechanisms such as thermal motion are amplified due to the increased sensitivity of the multi-ion interference to phase deviations~\cite{obvsil2019multipath}. At the same time, increasing the number of ions while preserving a linear configuration imposes an upper limit on the axial trapping frequencies  and thus, on the position uncertainties at given laser cooling temperature limits. In other words, larger ensembles of emitters lead to smaller angular size for the interference fringes, which makes the system all the more sensitive to these fluctuations. For example, three-dimensional ellipsoid crystals enable the coupling of many ions to the same optical mode~\cite{obvsil2018nonclassical, kovalenko2023emergence} and could improve detection efficiency, yet micro-motion severely affects phase-coherent phenomena in  these configurations. Since atomic motion blurs the interference from the coherent scattering, one then expects to recover the chaotic light statistics from phase-randomized emission. The suppression of coherent effects could be addressed with close-to-ground-state cooling techniques used for long ion chains in linear Paul traps~\cite{lechner2016electromagnetically,feng2020efficient,ejtemaee20173d}. Alternative platforms such as strings of neutral atoms also offer a favorable scaling for coherent phenomena due to the quasi-regular positions of the particles~\cite{tamura2020phase}.

Finally, beyond two-photon correlations, a natural question is how three- or higher-photon correlations can be manipulated using the interference between spontaneous emission and coherent scattering from multiple emitters. 
In particular, combining mechanisms such as interactions between the emitters~\cite{Liang2018,Cantu2020}, (projective) measurements~\cite{Zanthier2006,Wiegner2015,Bojer2022}, or frequency filtering~\cite{LopezCarreno2018, ZubizarretaCasalengua2020, Hanschke2020, Phillips2020, Masters2023}, along with the interference effect discussed in the present work~\cite{Bojer2024}, could lead to samples with different higher-order photon statistics in different directions.

\begin{acknowledgments}
{\it Acknowledgments---}The authors thank Robin Kaiser, Mathilde Hugbart, Joachim von Zanthier, and Manuel Bojer for fruitful discussions. A.C. and R.B. acknowledge the financial support of the São Paulo Research Foundation (FAPESP) (Grants No. 2022/06449-9, 2023/07463-8, 2018/15554-5, 2019/13143-0, 2023/07100-2, 2022/00209-6, and 2023/03300-7), from the Brazilian CNPq (Conselho Nacional de Desenvolvimento Científico e Tecnológico), Grant No. 313632/2023-5. A.~K. acknowledges the support of the Czech Science Foundation under the project GA21-13265X. D.~T., T.~P., and O.~Č. were supported by the national funding from the MEYS under the project CZ.02.01.01/00/22 008/0004649. L. S. is grateful for national funding from the MEYS under grant agreement No. 731473 and from the QUANTERA ERA-NET cofund in quantum technologies implemented within the European Union’s Horizon 2020 Programme (project PACE-IN, 8C20004). T. M. P. and O. Č. acknowledge the project 23FUN03 HIOC, which has received funding from the European Partnership on Metrology, co-financed from the European Union’s Horizon Europe Research and Innovation Programme and by the Participating States. K. S. acknowledges the project IGA-PrF-2024-008 of Palacky University Olomouc.
\end{acknowledgments}

\bibliography{Biblio}

\onecolumngrid
\newpage



\setcounter{equation}{0}
\setcounter{figure}{0}
\setcounter{table}{0}
\setcounter{page}{1}
\renewcommand{\theequation}{S\arabic{equation}}
\renewcommand{\thefigure}{S\arabic{figure}}

\begin{center}
{\large{ {\bf Supplemental Material for: \\ Coherent Control of Photon Correlations in Trapped Ion Crystals}}}

\vskip0.5\baselineskip{K. Singh,$^{1}$ A. Cidrim,$^{2}$ A. Kovalenko,$^{1}$ T. Pham,$^{1}$ O. Číp,$^{1}$ L. Slodi\v{c}ka,$^{1}$ and R. Bachelard$^{2}$}

\vskip0.5\baselineskip{ {\it $^{1}$Department of Optics, Palack\'{y} University, 17. listopadu 12, 771 46 Olomouc, Czech Republic\\
$^{2}$Departamento de F\'isica, Universidade Federal de S\~ao Carlos,\\ Rodovia Washington Lu\'is, km 235—SP-310, 13565-905 S\~ao Carlos, SP, Brazil
}}

\end{center}

\appendix

\setcounter{equation}{0}
\setcounter{figure}{0}
\setcounter{table}{0}
\setcounter{page}{1}

\twocolumngrid
\section{Experimental setup}

\begin{figure}[t!]
\centering	
\includegraphics[width=\columnwidth]{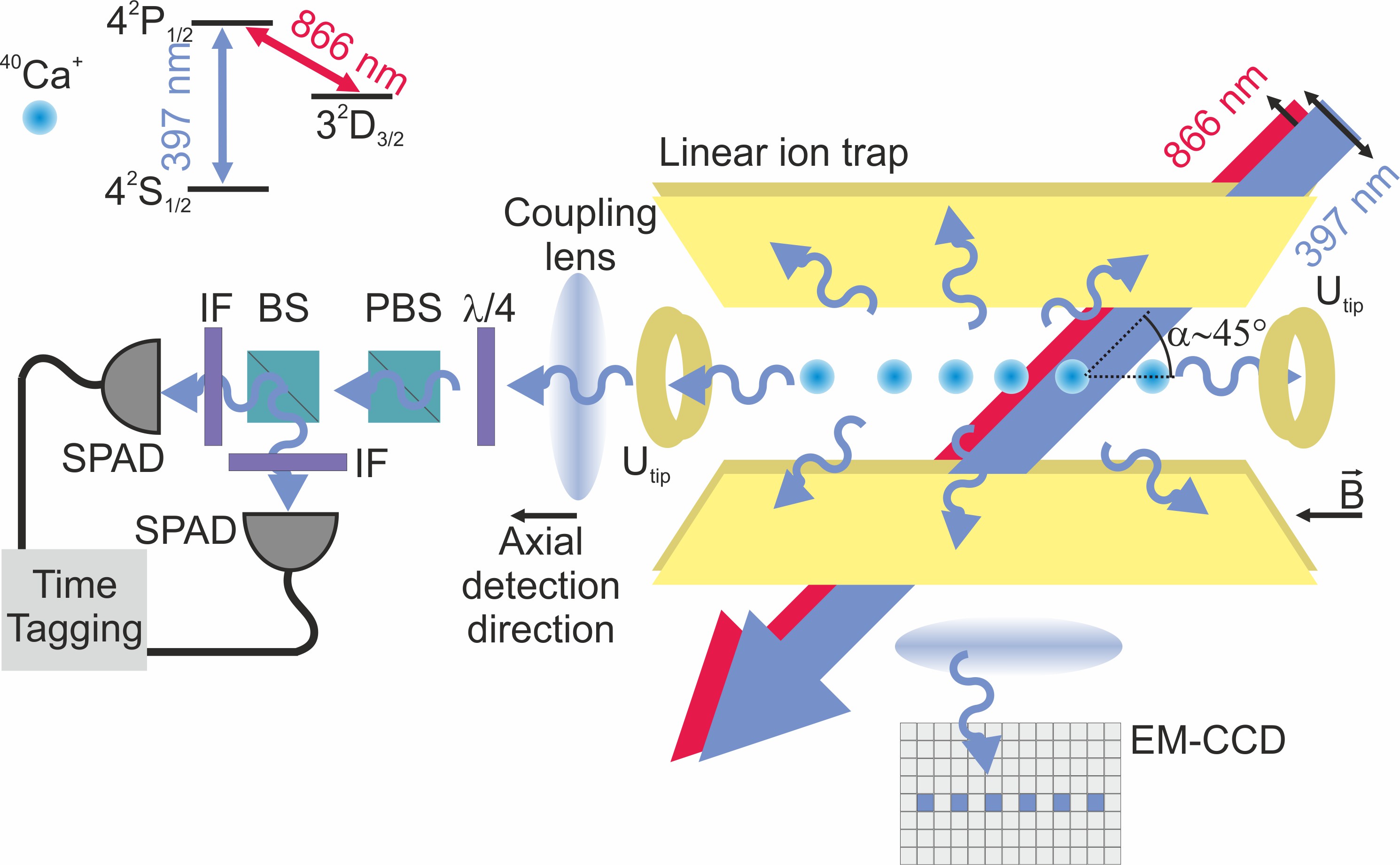}
\caption{Experimental scheme for observation of photon correlations on light from long ion crystals in a coherent scattering regime and relevant energy level scheme of $^{40}$Ca${^+}$ ion.}
\label{fig:experimentSupp}
\end{figure}

The measurements of the statistics of the light scattered by the ion crystals in free space are realized using the setup depicted in Fig.~\ref{fig:experimentSupp}. The strings of $^{40}$Ca${^+}$ ions are trapped in a linear Paul trap setup with an ion number ranging from $N=2$ to $N=18$. The ions are axially confined by a tunable static voltage $U_{\rm tip}= 50-1200$~V applied to the two hollow axial electrodes, which allows for a precise tuning of the ion positions in a single harmonic potential. The corresponding trapping axial frequency is mapped to the static voltage as $\omega_z = \kappa\sqrt{U_{\rm tip}}$, with $\kappa\approx (2\pi) 32\times 10^3$. Thus, the highest $U_{\rm tip}= 1200~{\rm\,V}$ corresponds to an axial motional frequency $\omega_z \approx (2\pi) 1.1$~MHz. A radio-frequency electric field applied to radial electrodes at $\omega_{\rm RF}=(2\pi) 29.9$~MHz and power of 4~W results in a confinement with secular frequencies of the radial center-of-mass mode~$\omega_x \approx \omega_y \approx (2\pi) 2.0$~MHz. The ions are continuously Doppler cooled using a 397~nm laser beam red-detuned from the $4^2{\rm S}_{1/2}\leftrightarrow 4^2{\rm P}_{1/2}$ transition by $\Delta_{397}\approx (2\pi) 12$~MHz. The angle between the 397~nm beam and the ion crystal axis has been estimated from the fit of the measured interference patterns to $\alpha = (45.1 \pm 0.1)^{\circ}$. The population of the metastable $3^2$D$_{3/2}$ manifold is continuously reshuffled back to the cooling transition using 866~nm laser light. A static magnetic field with a magnitude $|\vec{B}|=3.3$~Gauss is applied along the observation axial direction to lift the degeneracy of Zeeman states. Both excitation laser polarizations are set to linear, with an oscillation direction in the plane containing the excitation beam and the vector of the applied magnetic field. The addressed electronic transitions correspond to circular dipoles, observed in the axial detection direction along the static magnetic field.

The light scattered by the ions is collected with a lens of focal length 100~mm along the axial trapping direction. The observation solid angle corresponds to an effective numerical aperture NA~$\approx 0.07$, which is limited by the aperture of the axial trapping electrode positioned at 2.25~mm from the trap center. Note that for this numerical aperture and circular dipole orientation, this corresponds to an effective coupling to the detection mode of about $1.6\times 10^{-3}$, comparable to the $\beta$-parameter describing the atom-waveguide coupling reported in Refs.~\cite{Hinney2021,Cordier2023}. A combination of quarter waveplate ($\lambda/4$) and polarization beam-splitter~(PBS) is set to transmit photons from the selected $\sigma_{+}$~transition $4{\rm S}_{1/2} (m=-1/2) \leftrightarrow 4{\rm P}_{1/2} (m=+1/2)$. The populations of all other S, P, and D-levels add to the effective probability of ions being dark for the photon measurement. The populations were estimated using a fit of 8-level optical Bloch equations to the measured fluorescence spectra as a function of the 866 nm laser detuning, resulting in $P(4{\rm S}_{1/2} (m=-1/2)) \approx 0.57$,  $P(4{\rm S}_{1/2} (m=+1/2)) \approx 0.22$,  $P(4{\rm P}_{1/2} (m=-1/2)) \approx 0.04$, $P(4{\rm P}_{1/2} (m=+1/2)) \approx 0.04$, $P(3{\rm D}_{1/2} (m=-3/2)) \approx 0.03$, $P(3{\rm D}_{1/2} (m=-1/2)) \approx 0.04$, $P(3{\rm D}_{1/2} (m=+1/2)) \approx 0.04$, $P(3{\rm D}_{1/2} (m=+3/2)) \approx 0.02$. The collimated output light mode then propagates until a Hanbury-Brown and Twiss detection setup consisting of a bulk 50:50 beam splitter~(BS) and two single-photon counting modules~(SPADs) in each of its output ports. Both detection channels include a focusing lens with a focal length of 50~mm and a narrow-band optical interference filter~(IF) to optimize the coupling to the SPAD active area and suppress the spurious background counts, respectively. For two-ion crystals, the average count rate of photons at each of the two detectors was 1900~counts/s and 4300~counts/s for the measurement with $s= 0.6$ and 1.2, respectively. The light collected in the radial trapping direction is focused into an EM-CCD camera to determine precisely the number of trapped ions, as well as the spatial extent of the crystal before and after each measurement of the photon correlations.

The residual position uncertainty of ions, as well as the ratio of the coherent scattering to spontaneous emission, set the limit on the phase coherence between different ions~\cite{eschner2001light, obvsil2019multipath, wolf2016visibility}. The excitation laser settings, including the 397~nm and 866~nm lasers, detunings and saturation parameters, affect both these decoherence mechanisms simultaneously, so they have been directly optimized to maximize the interference contrast and were kept the same for all realizations. The presented saturation parameter of the excitation laser beam at 397~nm is defined as $s=2 \Omega^2/(4\Delta^2+\Gamma^2)$, with $\Omega$ the resonant Rabi frequency and $\Gamma=(2\pi) 21.2$~MHz the natural linewidth of the $4^2{\rm S}_{1/2}\leftrightarrow 4^2{\rm P}_{1/2}$ transition. The provided saturation parameters and detuning from transition resonance are estimated from dark resonance spectroscopy measurements on a single ion positioned in the trap center in the same excitation conditions. It decreases slightly for ions away from the center due to the Gaussian intensity profile of the 397~nm excitation beam --- the maximal decrease is about 10~\% for strings of up to 9 ions. However, for the crystal with $N=18$~ions, with a total length of $94\,\mu$m, for the ions positioned at the ends this decrease reaches $\approx 44$~\%, which represents a technical limitation to investigate the autocorrelation fluctuations for larger systems in our setup.

The ion chain setup is operated in the linear regime, far from the structural phase transition~\cite{schiffer1993phase}, so its spatial structure is fully determined by the applied axial DC-voltages. In particular, the small axial motional frequency drift estimated to $0.4 \pm 0.1$~Hz/min at $\omega_{\rm z}=(2\pi) 1.1$~MHz has a negligible effect on the stability of the generated interference patterns.

\section{Second-order auto-correlation function $g^{(2)}(0)$ for $N$ non-interacting atoms\label{app:g20}}

The electric field scattered by an ensemble of two-level atoms and along an observation direction $\hat n$ in the far-field reads
\begin{equation}\label{Eq:FarField}
    \hat E^{+} = E_0 \sum_{a=1}^{N} e^{-ik\hat n\cdot \mathbf{r}_a} \hat\sigma^{-}_a,
\end{equation}
with $\mathbf{r}_a$ the position vector of atom $a$, $\hat\sigma^{\mp}_a$ the Pauli lowering/raising spin operators, and $E_0$ a normalization prefactor. Without loss of generality, we set $E_0=1$, which normalizes the electric field intensity to unity for a single atom. We assume that the atoms are far apart from each other, so that light-induced dipole-dipole interactions between them can be disregarded. Therefore, the steady state of the system is separable and can be expressed as a product state as follows
\begin{equation}\label{Eq:SeparableState}  \hat\rho=\bigotimes_{a=1}^{N}\hat\rho_a,
\end{equation}
where $\hat\rho_a$ is the single-particle density matrix. As discussed in the main text, this assumption is justified for the inter-ion distances in our Paul trap (typically of a few wavelengths of the two-level transition). 

With Eq.~\eqref{Eq:FarField} we can thus calculate the intensity of the electric field scattered by the atoms as follows
\begin{equation}\label{Eq:IntensityDefition}
\begin{aligned}
I&=\langle \hat E^{-} \hat E^+\rangle=\sum_{ab}\Tr{e^{ik\hat n\cdot \mathbf{r}_a}\hat\sigma_a^{+}e^{-ik\hat n\cdot \mathbf{r}_b}\hat\sigma^{-}_b\hat\rho}.
\end{aligned}
\end{equation}
The trace operation can be further evaluated assuming the separable atomic state of Eq.~\eqref{Eq:SeparableState}, simplifying the expression for the intensity to
\begin{equation}
\begin{aligned}\label{Eq:ISeparable}
&I=\sum_{a}n_a+\Big|\sum_{a}\beta_a\Big|^2-\sum_{a}|\beta_a|^2,
\end{aligned}
\end{equation}
where we have defined the single-atom population as
\begin{equation}\label{Eq:PopulationDefinition}
    n_a \equiv \Tr{\hat\sigma_a^{+}\hat\sigma^{-}_a\hat\rho_a}
\end{equation}
and the corresponding atomic coherence (with an additional observation phase) as
\begin{equation}\label{Eq:CoherenceDefinition}
    \beta_a \equiv e^{-ik\hat n\cdot \mathbf{r}_a}\Tr{\hat\sigma^{-}_a\hat\rho_a}\implies \beta_a^{*} \equiv e^{ik\hat n\cdot \mathbf{r}_a}\Tr{\hat\sigma^{+}_a\hat\rho_a}.
\end{equation}

The same steps can be taken to derive the unnormalized second-order correlation function
\begin{equation}\label{Eq:SecondOrderDefinition}
\begin{aligned}
G^{(2)}&(0)=\langle \hat E^{-} \hat E^{-}\hat E^+ \hat E^+\rangle,
\end{aligned}
\end{equation}
which, again for the separable state \eqref{Eq:SeparableState} and in terms of single-atom population and coherence, results in the expression 
\begin{equation}\label{Eq:G2Separable}
\begin{aligned}
G^{(2)}(0)=&2\Big(\sum_a n_a\Big)^2-2\sum_a n_a^2 
+4\Big(\sum_a n_a\Big)\Big(\Big|\sum_b\beta_b\Big|^2-\sum_b|\beta_b|^2\Big)\\
-&8\mathrm{Re}\Big\{\Big(\sum_a n_a\beta^*_a\Big)\Big(\sum_b \beta_b\Big)\Big\}+8\sum_a n_a |\beta_a|^2\\
+&\Big|\sum_a \beta_a\Big|^4-6\sum_a |\beta_a|^4 - 4\Big|\sum_a \beta_a\Big|^2\Big(\sum_b |\beta_b|^2\Big)\\
+&8\mathrm{Re}\Big\{\Big(\sum_a \beta_a\Big)\Big(\sum_b |\beta_b|^2\beta_b^*\Big)\Big\}+2\Big(\sum_a |\beta_a|^2\Big)^2\\
-&2\mathrm{Re}\Big\{\Big(\sum_a \beta_a\Big)^2\Big(\sum_b (\beta^*_b)^2\Big)\Big\}+\Big|\sum_a \beta^2_a\Big|^2.
\end{aligned}
\end{equation}

Let us now consider that the separable steady-state \eqref{Eq:SeparableState} is created by resonantly illuminating the two-level atoms with a laser with wave-vector $\mathbf{k}_L$, Rabi frequency $\Omega$ and detuning $\Delta$. Therefore, the resulting single-atom density matrix is given by
\begin{equation}
\hat\rho_a=\rho^{(a)}_{ee}|e\rangle\langle e|+\rho^{(a)}_{eg}|e\rangle\langle g|+\rho^{(a)}_{ge}|g\rangle\langle e|+\rho^{(a)}_{gg}|g\rangle\langle g|,
\end{equation}
with 
\begin{equation}
\begin{aligned}
\rho^{(a)}_{ge}&=\Big(\rho^{(a)}_{eg}\Big)^*=-i\frac{e^{-i\mathbf{k}_L.\mathbf{r}_a}}{1+s}\sqrt{\frac{s}{2}},\\ \rho^{(a)}_{ee}&=\frac{s}{2(1+s)},\,\,\rho^{(a)}_{gg}=\frac{2+s}{2(1+s)},
\end{aligned}    
\end{equation}
where $s\equiv 2\Omega^2/(\Gamma^2+4\Delta^2)$ is the saturation parameter, as introduced in the main text. Thus, Eqs.~\eqref{Eq:PopulationDefinition} and \eqref{Eq:CoherenceDefinition} can be rewritten in terms of the saturation parameter and reduce to 
\begin{equation}\label{Eq:PopulationSaturation}
    n_a = \frac{s}{2(1+s)},
\end{equation}
and
\begin{equation}\label{Eq:CoherenceSaturation}
    \beta_a = -i \frac{e^{-i(k\hat n-\mathbf{k}_L)\cdot\mathbf{r}_a}}{1+s}\sqrt{\frac{s}{2}}.
\end{equation}

Finally, the normalized auto-correlation function as defined by $g^{(2)}(0)\equiv G^{(2)}(0)/I^2$ can be directly computed from Eqs.~\eqref{Eq:ISeparable}, \eqref{Eq:G2Separable}, \eqref{Eq:PopulationSaturation}, and \eqref{Eq:CoherenceSaturation}, resulting in the expression~\eqref{eq:g20} shown in the main text.

\subsection{Measured temporal second-order correlation functions}

\begin{figure*}[t!]
\centering	
\includegraphics[width= \textwidth]{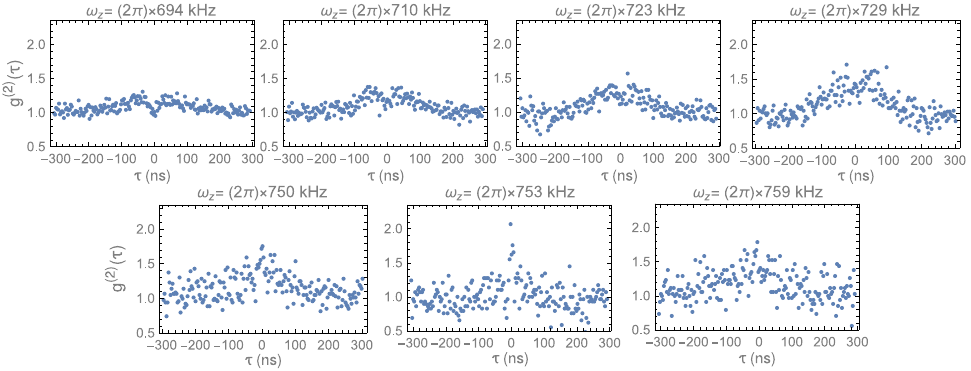}
\caption{Measured temporal profiles of second-order correlation function $g^{(2)}(\tau)$ for different axial frequencies of the linear crystal with $N=4$ ions. The temporal dependence close to a zero time delay reveals the transition from anti-bunched [i.e., $g^{(2)}(0) < g^{(2)}(|\tau|\approx 50\,{\rm ns})$] to bunched photon statistics for close to constructive to destructive interference of the scattered photons at $\approx 700$~kHz and $\approx 750$~kHz, respectively. The higher noise in the measurements corresponding to the close-to-destructive interference results from the lower detectable photon rate.}
\label{fig:data4ions}
\end{figure*}

We here provide the experimental second-order correlation functions $g^{(2)}(\tau)$ within the temporal range of the observed coherence timescale ($\sim 300$ns). The values of $g^{(2)}(0)$ presented in the main text are extracted from the present $g^{(2)}(\tau)$. 

Each curve labeled with the axial motional frequency in Fig.~\ref{fig:data4ions} relates to a data point in Fig.~2 of the main text for the case of a 4-ion crystal.

\begin{figure*}[h!]
\centering	
\includegraphics[width=\textwidth]{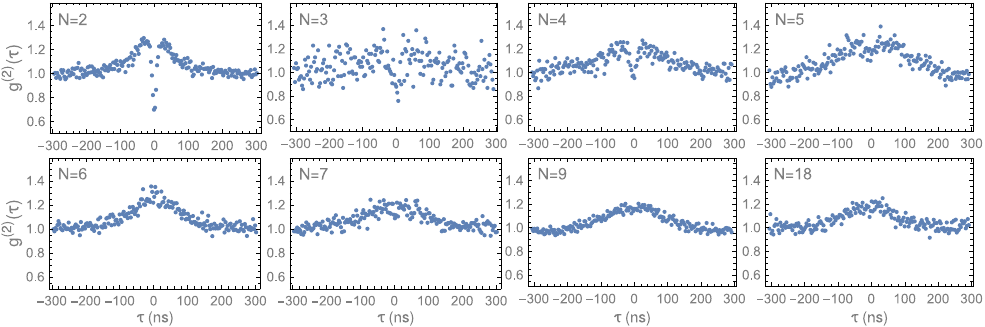}
\caption{Measured second-order correlation functions for different numbers of ions~$N$ at constructive interference points and saturation parameter $s=0.6$. The profiles confirm the observation of the transition from sub-Poissonian and anti-bunched photon statistics to super-Poissonian light.}
\label{fig:lowS}
\end{figure*}

\begin{figure*}[h!]
\centering	
\includegraphics[width=\textwidth]{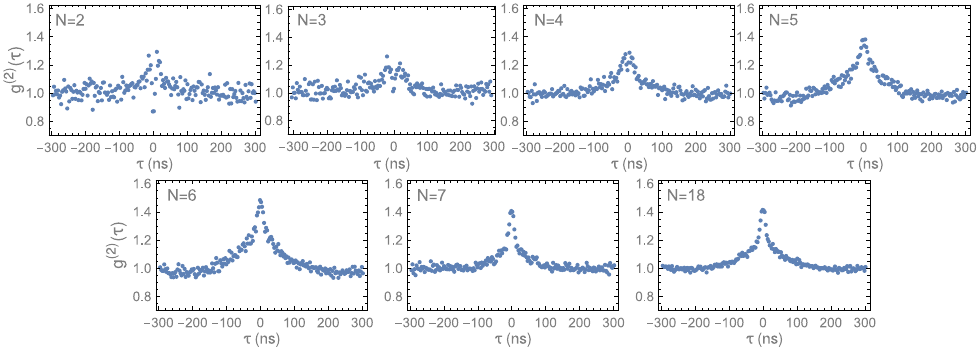}
\caption{Measured second-order correlation functions for different numbers of ions~$N$ at constructive interference points and saturation parameter $s=1.2$. The measured profiles confirm the observation of the transition from sub-Poissonian (for $N=2$) and anti-bunched photon statistics to unambiguously bunched and super-Poissonian photon statistics. Here, the fast dynamics of the photon emission at a high saturation parameter suppresses the contrast of the resulting narrow temporal features in $g^{(2)}(\tau)$ for the given jitter of the employed single photon counting modules of around $1$~ns.}
\label{fig:highS}
\end{figure*}

Data points from Fig.~3 in the main text for different number of ions in the linear crystal can be paired with the corresponding $g^{(2)}(\tau)$ in Figs.~\ref{fig:lowS} and~\ref{fig:highS}, for saturation parameter (of the excitation laser beam at $397$~nm)  $s=0.6$ and $1.2$, respectively.

In addition to the directly recognizable sub-Poissonian and super-Poissonian character of the measured $g^{(2)}(\tau)$ at zero-time delay, the temporal profiles allow for the monitoring of the evolution of the bunching [i.e., $g^{(2)}(0) > g^{(2)}(|\tau| > 0)$ for any $\tau$] and anti-bunching [$g^{(2)}(0) < g^{(2)}(|\tau| > 0)$ for a given $\tau$] of the light scattered coherently by the linear ion chain~\cite{Loudon:book}. The presented temporal photon correlations $g^{(2)}(\tau)$ are time-binned with a resolution of $3$~ns.

\end{document}